\documentclass[sigconf]{acmart}

\usepackage[normalem]{ulem}

\usepackage[ruled,vlined,linesnumbered]{algorithm2e}
\usepackage{multirow}
\usepackage{listings}

\usepackage{color}
\usepackage{graphicx}

\DeclareRobustCommand*\cal{\relax\mathcal}

\copyrightyear{2020} 
\acmYear{2020} 
\setcopyright{acmcopyright}
\acmConference[ICSE '20]{42nd International Conference on Software Engineering}{May 23--29, 2020}{Seoul, Republic of Korea}
\acmBooktitle{42nd International Conference on Software Engineering (ICSE '20), May 23--29, 2020, Seoul, Republic of Korea}
\acmPrice{15.00}
\acmDOI{10.1145/3377811.3380334}
\acmISBN{978-1-4503-7121-6/20/05}
\begin{document}
\title{sFuzz: An Efficient Adaptive Fuzzer for Solidity Smart Contracts}

\author{Tai D. Nguyen}
\affiliation{%
  \institution{Singapore Management University, Singapore}
  }
\email{dtnguyen.2019@smu.edu.sg}

\author{Long H. Pham}
\affiliation{%
  \institution{Singapore Management University, Singapore}
  }
\email{longph1989@gmail.com}

\author{Jun Sun}
\affiliation{%
  \institution{Singapore Management University, Singapore}
  }
\email{sunjunhqq@gmail.com}

\author{Yun Lin}
\affiliation{%
  \institution{National University of Singapore, Singapore}
  }
\email{llmhyy@gmail.com}

\author{Quang Tran Minh}
\affiliation{%
  \institution{Ho Chi Minh City University of Technology, Vietnam}
  }
\email{quangtran@hcmut.edu.vn}


\begin{abstract}
Smart contracts are Turing-complete programs that execute on the infrastructure of the blockchain, which often manage valuable digital assets. Solidity is one of the most popular programming languages for writing smart contracts on the Ethereum platform. Like traditional programs, smart contracts may contain vulnerabilities. Unlike traditional programs, smart contracts cannot be easily patched once they are deployed. It is thus important that smart contracts are tested thoroughly before deployment. In this work, we present an adaptive fuzzer for smart contracts on the Ethereum platform called sFuzz. Compared to existing Solidity fuzzers, sFuzz combines the strategy in the AFL fuzzer and an efficient lightweight multi-objective adaptive strategy targeting those hard-to-cover branches. 
sFuzz has been applied to more than 
4 thousand 
smart contracts and the experimental results show that (1) sFuzz is efficient, e.g., two orders of magnitude faster than state-of-the-art tools; (2) sFuzz is effective in achieving high code coverage and discovering vulnerabilities; and (3) the different fuzzing strategies in sFuzz complement each other.
\end{abstract}

\maketitle

\section{Introduction}
Nowadays, smart contracts~\cite{clack2016smart,szabo1997formalizing} are implemented as Turing-complete programs that execute on the infrastructure of the block-chain~\cite{zheng2016blockchain}. It provides a framework that potentially allows any program (equivalently, contract) to be executed in an autonomous, distributed, and trusted way. Smart contracts thus have the potential to revolutionize many industries. Popular applications of smart contracts include crowd fundraising, online gambling and so on. Ethereum~\cite{ewp,eyp} is the first to introduce the functionality of smart contracts. Based on the Ethereum platform, Solidity is the most popular programming language for smart contracts~\cite{solidity}.

Like traditional C or Java programs, smart contracts may contain vulnerabilities. Unlike traditional programs, smart contracts cannot be modified easily once they are deployed on the blockchain~\cite{marino2016setting}. As a result, a vulnerability renders the smart contract forever vulnerable, which significantly magnifies the problem. In recent years, there has been an increasing number of news reports on attacks
which exploit security vulnerabilities in Ethereum smart contracts. One particularly noticeable example is the DAO attack~\cite{dao}, i.e., an attacker stole more than 3.5 million Ether (which is equivalent to about \$45 million USD at the time) exploiting a vulnerability in the DAO contract. To fix the vulnerability, a hard fork was launched which was not only expensive but also caused much controversy~\cite{dao}.

It is thus desirable to develop tools for validating smart contracts to identify vulnerabilities, ideally before they are deployed. Among the range of complementary techniques for validating smart contracts, we focus on automatic testing of smart contracts in this work as testing is often the least expensive and thus the most applicable. To automatically test smart contracts, we must solve the following three problems:
\begin{itemize}
    \item the test automation problem (i.e., how to run test cases),
    \item the test generation problem (i.e., what to test),
    \item and the oracle problem (i.e., what are vulnerabilities).
\end{itemize}
In the literature, several approaches have been developed for automatic testing smart contracts, each of which answers these three problems in slightly different ways. For instance, ContractFuzzer~\cite{Jiang:2018:CFS:3238147.3238177} builds a network with pre-deployed contracts and generates transactions to run smart contracts, generates test cases based on a set of predefined parameter values and targets a set of oracles specific for smart contracts. Oyente~\cite{luu2016making} runs smart contracts symbolically through symbolic execution, generates test cases for covering different program paths in single functions through constraint solving, and supports multiple oracles to identify 4 kinds of vulnerabilities. 
teEther~\cite{DBLP:conf/uss/KruppR18} similarly applies symbolic execution to generate test cases covering program paths, and focuses on oracles which are related to financial transactions.

In this work, we propose a fully automatic testing engine for smart contracts running on Ethereum called sFuzz. sFuzz is inspired by AFL~\cite{afl}, a well-known fuzzer for C programs, i.e., sFuzz is a feedback-guided fuzzing engine and is inexpensive to apply. sFuzz complements existing testing engines based on symbolic execution like Oyente and teEther, as it is known that fuzzing and symbolic execution are complementary~\cite{wang2018towards,yun2018qsym}.
While AFL-based fuzzing is often effective, it has its limitation as well, i.e., it is often expensive in covering branches guarded with strict conditions. To tackle the problem, sFuzz integrates AFL-based fuzzing with an efficient lightweight adaptive strategy for selecting seeds. 
Although inspired by search-based software testing~\cite{harman2010theoretical,mcminn2004search}, the latter distinguishes itself by having a lightweight objective function (designed considering characteristics of Solidity programs) as well as a novel multi-objective optimization strategy. 

sFuzz is built based on Aleth~\cite{aleth} (i.e., an Ethereum VM written in C++), has a system architecture similar to AFL, and is extensible to different Ethereum VMs and oracles as well as fuzzing strategies. sFuzz has been systematically applied to a set of more than 
4 thousand smart contracts. The experimental results show that sFuzz is on average more than two orders of magnitudes faster than ContractFuzzer, covers more branches and reveals many more vulnerabilities. A comparison between sFuzz and Oyente shows that they are complementary. 
Furthermore, experiments with prolonged fuzzing time show that the adaptive strategy improves code coverage. sFuzz is available online and has been adopted by multiple companies.

The remainder of the paper is organized as follows. Section~\ref{example} illustrates how sFuzz works through examples. Section~\ref{sec:fuzz/smart/contracts} presents the details of the approach. Section~\ref{sec:implementation} shows implementation details of sFuzz. Section~\ref{sec:evaluation} reports evaluation results. Section~\ref{sec:related} reviews related work and 
concludes.
 \label{intro}
\section{Illustrative Examples}
In this section, we show how sFuzz works step-by-step through two illustrative examples. Note that Solidity source codes for both examples are shown for simplicity. sFuzz requires only the EVM (i.e., Ethereum Virtual Machine) bytecode~\cite{ewp, eyp} to fuzz smart contracts.

Given a smart contract, sFuzz automatically configures a block-chain network, deploys the smart contract, and generates multiple transactions each of which calls a function in the contract. The transactions are then executed with an EVM enriched with a set of oracles for identifying vulnerabilities. sFuzz monitors the execution of the transactions to collect certain feedback, e.g., whether a certain branch has been covered and how far the branch is covered. Whenever a vulnerability is revealed, the transactions and the network configuration (i.e., a test case) are saved and reported to the user later on. Otherwise, some of the test cases are selected as $seeds$ based on feedback collected during the transaction execution according to certain seed selection criteria. Afterwards, the $seeds$ are mutated to generate the next generation of test cases. This process repeats until a time out occurs.



\definecolor{verylightgray}{rgb}{.97,.97,.97}

\lstdefinelanguage{Solidity}{
	keywords=[1]{anonymous, assembly, assert, balance, break, call, callcode, case, catch, class, constant, continue, constructor, contract, debugger, default, delegatecall, delete, do, else, emit, event, experimental, export, external, false, finally, for, function, gas, if, implements, import, in, indexed, instanceof, interface, internal, is, length, library, log0, log1, log2, log3, log4, memory, modifier, new, payable, pragma, private, protected, public, pure, push, require, return, returns, revert, selfdestruct, send, solidity, storage, struct, suicide, super, switch, then, this, throw, transfer, true, try, typeof, using, value, view, while, with, addmod, ecrecover, keccak256, mulmod, ripemd160, sha256, sha3}, 
	keywordstyle=[1]\color{blue}\bfseries,
	keywords=[2]{address, bool, byte, bytes, bytes1, bytes2, bytes3, bytes4, bytes5, bytes6, bytes7, bytes8, bytes9, bytes10, bytes11, bytes12, bytes13, bytes14, bytes15, bytes16, bytes17, bytes18, bytes19, bytes20, bytes21, bytes22, bytes23, bytes24, bytes25, bytes26, bytes27, bytes28, bytes29, bytes30, bytes31, bytes32, enum, int, int8, int16, int24, int32, int40, int48, int56, int64, int72, int80, int88, int96, int104, int112, int120, int128, int136, int144, int152, int160, int168, int176, int184, int192, int200, int208, int216, int224, int232, int240, int248, int256, mapping, string, uint, uint8, uint16, uint24, uint32, uint40, uint48, uint56, uint64, uint72, uint80, uint88, uint96, uint104, uint112, uint120, uint128, uint136, uint144, uint152, uint160, uint168, uint176, uint184, uint192, uint200, uint208, uint216, uint224, uint232, uint240, uint248, uint256, var, void, ether, finney, szabo, wei, days, hours, minutes, seconds, weeks, years},	
	keywordstyle=[2]\color{teal}\bfseries,
	keywords=[3]{block, blockhash, coinbase, difficulty, gaslimit, number, timestamp, msg, data, gas, sender, sig, value, now, tx, gasprice, origin},	
	keywordstyle=[3]\color{violet}\bfseries,
	identifierstyle=\color{black},
	sensitive=false,
	comment=[l]{//},
	morecomment=[s]{/*}{*/},
	commentstyle=\color{gray}\ttfamily,
	stringstyle=\color{red}\ttfamily,
	morestring=[b]',
	morestring=[b]"
}

\lstset{
	language=Solidity,
	backgroundcolor=\color{verylightgray},
	extendedchars=true,
	basicstyle=\footnotesize\ttfamily,
	showstringspaces=false,
	showspaces=false,
	numbers=left,
	numberstyle=\footnotesize,
	numbersep=9pt,
	tabsize=2,
	breaklines=true,
	showtabs=false,
	captionpos=b
}

\begin{figure}[t]
\centering
\begin{minipage}[h]{0.95\linewidth}
\begin{lstlisting}
pragma solidity ^0.4.20;
contract opposite_game {
  string public question;
  address questionSender;
  bytes32 responseHash;

  function Try(
    string _response ) external payable {
    if(responseHash == keccak256(_response) &&
        msg.value == 100 finney) {
      msg.sender.send(this.balance); } }
    
  function start_quiz_game(
    string _question, string _answer) public payable {
    if(responseHash==0x0) {
      responseHash = keccak256(_answer);
      question = _question;
      questionSender = msg.sender; } }
    
  function() public payable {} }
\end{lstlisting}
\end{minipage}
\caption{An example with single objective function}
\label{examplesol}
\end{figure}

In the following, we describe how sFuzz works using the contract shown in Figure~\ref{examplesol}. The contract implements a simple quiz game. The contract is based on contract $opposite\_game$\footnote{address: 0x467532e79222670a2044c9b168bcbaa33b390ef5} with minor modification for simplicity. A quiz can be created by calling function $start\_quiz\_game$. The response is hashed and then saved in the $responseHash$ variable. The user then calls the $try$ function with their answer as the argument and pays a fee of $100~finney$ (which is a unit of the token) for each try. If the answer is correct, a reward is sent to the user.

This contract suffers from a vulnerability known as \emph{Gasless Send} when line 11 is executed and a costly \emph{fallback function} is called. That is, when function $send()$ at line 11 is executed, if the receiver is a contract, its fallback function is executed automatically. Because function $send()$ only forwards 2300 units of gas (i.e., price to pay for executing the function), an \emph{out-of-gas} exception is thrown if the fallback function is costly (e.g., costs more than 2300 units of gas). In this case, the $send()$ function simply returns $false$ and because the returned value is not checked and handled accordingly, the owners of the contract can keep the reward for themselves.

To expose this vulnerability, first a network is configured with several addresses and associated balances.
This contract is then deployed at one of the addresses. In addition, an attacker contract with a costly fallback function is deployed automatically. To expose the vulnerability, a test case (i.e., a sequence of transactions) with such a network configuration must first call function $start\_quiz\_game$ and then function $Try$ with parameters such that all 2 conditions in function $Try$ at line 9 and 10 are satisfied. The condition at line 9 is satisfied with a test case that sets all the parameters and contract variables to the default value of $0$. Note that $responseHash$ is set to $keccak256(\_answer)$ at line 16 and is compared to $keccak256(\_response)$ at line 9. However, generating a test case which satisfies the second condition by randomly generated test values is highly unlikely. The variable $msg.value$ has a size of 32 bytes and thus we have only $\frac{1}{2^{256}}$ probability to generate the value 100 (if we generate random values with a uniform distribution among all possible values). Existing fuzzing strategy in AFL is ineffective in this case as well, i.e., AFL selects test cases that cover new branches as $seeds$. Since all test cases generated through mutation are unlikely to cover the then-branch at line 10, they are equally `bad' according to the AFL seed selection strategy.

sFuzz complements AFL's seed selection strategy with an adaptive strategy that prioritizes the seeds according to a quantitative measure (i.e., a distance) on how far a seed is from covering any just-missed branch.
For this example, the distance for covering the just-missed branch (i.e., the then-branch) is computed as: $|msg.value - 100| + 1$, based on the value of $msg.value$ when the branch at line 10 is reached in the test case. Intuitively, the smaller the distance is, the closer the test case is to cover the branch (i.e., with a $msg.value$ closer to $100$). In particular, when $msg.value$ is exactly 100, the distance value reaches the minimum value of 1.
Based on this measurement, sFuzz iteratively selects $seeds$ which gradually gets closer and closer to satisfying the condition at line 10. In our experiment, after 140 generations, sFuzz generates a test case which covers the branch, and reveals the vulnerability.

The above example shows a simplistic situation where there is only one just-missed branch. In general, there may be multiple just-missed branches and thus sFuzz measures a distance for each pair of test case and just-missed branch, i.e., how far is the branch from being covered by the test case. Then for each just-missed branch, sFuzz selects the test case with the minimum distance as the $seed$.
For instance, the contract in Figure~\ref{examplesol2} shows a function which performs some basic arithmetic operations. There are two different branches, i.e., the condition at line 5 for comparing $y$ with $110$ and the one at line 6 for comparing $y$ with $10010$. Assume that both then-branches are yet to be covered. Given any test case, sFuzz computes two distances, one for covering the first then-branch; 
and the other for covering the second then-branch. 
Given a set of test cases, sFuzz selects, for each of these two branches, a test case which has minimum distance as seed, to generate further test cases.
After repeating the process multiple times, sFuzz generates two test cases that cover the two then-branches. \emph{We remark that for this example, due to the non-linear computation at line 4, approaches based on symbolic execution like Oyente~\cite{luu2016making} and teEther~\cite{DBLP:conf/uss/KruppR18} are ineffective due to the limitation of underlying constraint solvers.}

\begin{figure}[t]
\centering
\begin{minipage}[h]{0.8\linewidth}
\begin{lstlisting}
pragma solidity ^0.4.20;
contract multiple_objective_function {
  function foo(int x) {
    int y = x*x + 10;
    if(y == 110) { ... }
    if(y == 10010) { ... } } }
\end{lstlisting}
\end{minipage}
\caption{An example with multiple objective functions}
\label{examplesol2}
\end{figure}

 \label{example}
\section{Fuzzing Smart Contracts}
\label{sec:fuzz/smart/contracts}
In this section, we define our problem and then present our approach in detail step-by-step.

%

\subsection{Problem Definition}
A smart contract $\cal S$ typically has a number of instance variables, a constructor and multiple functions, some of which are public. It can be equivalently viewed in the form of a control flow graph (CFG) ${\cal S} = (N, i, E)$ where $N$ is a finite set of control locations in the program; $i \in N$ is the initial control location, i.e., the start of the contract; and $E \subseteq N \times C \times N$ is a set of labeled edges, each of which is of the form $(n, c, n')$ where $c$ is either a condition (for conditional branches like if-then-else or while-loops) or a command (i.e., an assignment). Note that for simplicity, we define the smart contract as one single graph rather than defining one graph for each function and then connecting them through a call graph. A node in the graph is branching if and only if it has multiple child nodes and its outgoing edges are labeled with conditions. We refer to an outgoing edge of a branching node as a branch. \\

\noindent \emph{Test cases.} A test case for $\cal S$ is a pair $(\sigma_0, \Sigma)$ where $\sigma_0$ is a configuration of the blockchain network and $\Sigma$ is a sequence of transactions (i.e., function calls). The configuration $\sigma_0$ contains all information on the setup of the network which is relevant to the execution of the smart contract. Formally, $\sigma_0$ is a tuple $(b, ts, SA, SB, v)$ where $b$ is the current block number, $ts$ is the current block timestamp, $SA$ is a set of the addresses of the smart contracts (including the smart contract under test as well as other invoked contracts), $SB$ is a function which assigns an initial balance to each address and $v$ is the initial valuation of the persistent state.
$\Sigma = \langle m_0(\overrightarrow{p_0}), m_1(\overrightarrow{p_1}), \cdots \rangle$ is a sequence of public function calls of the smart contract under test, each of which has an optional sequence of concrete input parameters $\overrightarrow{p_i}$. Note that $m_0$ must be a call of the constructor.

The task of fuzzing a smart contract is thus to generate a set of test cases (a.k.a. test suite) according to certain testing criteria. The execution of a test case $t$ traverses through a path in the CFG $\cal S$, which visits a set of nodes and edges. For simplicity, we assume that one test execution covers one unique path (i.e., there is no non-determinism). Furthermore, a trace generated by $t$ is a sequence of pairs of the form $\langle (\sigma_0, n_0), (\sigma_1, n_1), \cdots \rangle$ where $(n_0, n_1, \cdots)$ is the sequence of nodes visited by $t$ and $\sigma_i$ is the configuration at the time of visiting node $n_i$ for all $i$.\\

\noindent \emph{Code Coverage.} Ideally, we aim to generate a test suite which reveals all vulnerabilities in the contract.
However, as we do not know where the vulnerabilities are, we must instead aim to achieve something more measurable. In this work, our answer is to focus on code coverage, in particular, branch coverage. We remark that our approach can be extended to support different coverage at the cost of additional code instrumentation. A branch in $\cal S$ is covered by a test suite if and only if there is a test case $t$ in the suite that visits the edge at least once. The branch coverage of a test suite is calculated as the percentage of the covered branches over the total number of branches. Note that identifying the total number of (feasible) branches statically in a smart contract is often infeasible for two reasons. First, some branches might be infeasible (i.e., there does not exist any test case that visits the branch) and knowing whether a branch is feasible or not is a hard problem. Second, EVM has a stack-based implementation which makes identifying all potentially feasible branches hard (as we will explain in more detail in Section~\ref{implementation}). \emph{Our problem is thus reduced to generate a test suite which maximizes the number of covered branches.} 

To achieve maximum code coverage, one way is to generate a large test suite (e.g., through random test generation). However, in practice, we often have limited resources (in terms of time or the number of computer processes) and thus our problem is refined as `\emph{to generate a test suite which maximizes the number of covered branches as efficiently as possible}'. Our solution to the problem is feedback-guided adaptive fuzzing.

Fuzzing is one of the most popular methods to create test cases~\cite{klees2018evaluating}. A feedback-guided fuzzing system (a.k.a.~fuzzer) takes a program under test and an initial test suite as input, monitors the execution of the test cases to obtain certain feedback, generates new test cases based on the existing ones in certain ways and then repeats the process until a stopping criteria is satisfied. We present details of our feedback-guided adaptive fuzzing process in Section~\ref{process}. \\

\noindent \emph{Oracles} The remaining problem is then how to tell whether a test case reveals a vulnerability. In this work, we adopt a set of oracles from previous approaches~\cite{Jiang:2018:CFS:3238147.3238177,luu2016making} including \emph{Gasless Send}, \emph{Exception Disorder}, \emph{Timestamp Dependency}, \emph{Block Number Dependency}, \emph{Dangerous DelegateCall}, \emph{Reentrancy}, \emph{Integer Overflow/Underflow}, and \emph{Freezing Ether}. We refer the readers to Section~\ref{implementation} for details.

\subsection{Feedback-Guided Adaptive Fuzzing} \label{process}
The general idea of feedback-guided fuzzing is to transform the test generation problem into an optimization problem and use some form of feedback as an \emph{objective function} in solving the optimization problem. Our fuzzing strategy is adaptive as we change the objective function adaptively based on the feedback. At the top level, sFuzz employs a genetic algorithm~\cite{genetic} which is inspired by the well-known AFL fuzzer to evolve the test suite in order to iteratively improve its branch coverage.

The overall workflow is shown in Algorithm~\ref{fig:algo}. Variable $suite$ is the test suite to be generated. It is initially empty. Whenever a test case covers a new branch, it is added into $suite$. Variable $seeds$ is a set of seed test cases, based on which new test cases are generated. First, we generate an initial test suite using function $initPopulation()$. The loop from line 3 to 6 then iteratively evolves the test suite. In particular, we add those test cases in $seeds$ which cover new branches (i.e., any branch which is not covered by test cases in $suite$) into $suite$ at line 4. At line 5, we filter the test cases in $seeds$ through function $fitToSurvive()$ so as to focus on those seeds which are more likely to lead to test cases covering new branches later. At line 6, function $crossoverMuatation()$ generates more test cases based on the test cases in $seeds$. The loop continues until a pre-set time out is triggered. While Algorithm~\ref{fig:algo} resembles the one in AFL, the differences are in the details of each function. In the following, we present each function in detail. \\

\begin{algorithm}[t]
\small
\SetAlgoVlined
let $suite$ be an empty test suite\;
let $seeds := initPopulation()$\;
\While{not time out} {
    add tests in $seeds$ which covers new branches into $suite$\;
    let $seeds := fitToSurvive(seeds)$\;
    let $seeds = crossoverMutation(seeds)$\;
}
return $suites$\;
\caption{The test generation algorithm}
\label{fig:algo}
\end{algorithm}


\noindent \emph{Generating Initial Population}
Function $initPopulation()$ generates an initial population containing multiple test cases. As mentioned above, to generate a test case, we need to generate an initial configuration $\sigma_0$ as well as a sequence of (public) function calls with concrete parameters. The initial configuration by default is as follows (in hexadecimal): $b = 0$, $ts = 0$, $SA = \{\texttt{0xf0}\}$,
$SB = \{\texttt{0xff00...}\}$
and $v$ is set using the declared initial value for each variable representing the persistent state. sFuzz additionally allows a user to customize the initial configuration, i.e., the user is allowed to provide an initial set of test cases.



Next, we generate multiple sequences of transactions, each of which is a function call with concrete parameters. For a contract with $n$ functions, we generate $n$ sequences. In each sequence, a different function is called once after the constructor is called. 
This makes sure that each function is tested at least once (i.e., function coverage is 100\%). 

For each function call, we generate a random value for each parameter based on its type. Note that if the parameter type has a fixed-length, e.g., of type $uint256$, this is straightforward. If the type does not have a fixed length (e.g., an array or a string), we first randomly generate a number (with a range from 0 to $bound$ where $bound$ is a bound on maximum length with a default value of 255) representing the number of elements in the parameter (e.g., number of characters) and then generate a corresponding number of element values.

\begin{figure*}[t]
    \centering
    \includegraphics[width=0.9\linewidth]{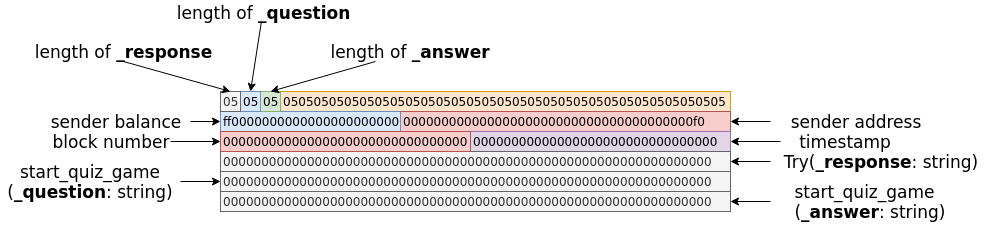}
    \caption{A generated test case}
    \label{fig:format}
\end{figure*}

Each test case is encoded in form of a bit vector. In the terminology of genetic algorithms, such bit vectors can be naturally regarded as \emph{chromosomes}. The size of the bit vector equals to the number of bits for encoding the configuration plus the number of bits encoding the function calls. Note that for each test case, we keep a list of function calls (which always includes the constructor in the contract)
and then encode each parameter value. If the parameter value is of variable-length, we use $\lceil \log bound \rceil$ (where $bound$ is a bound on the maximum length with a default value of 255) to encode the length of the parameter value. For example, given the contract shown in Figure~\ref{examplesol}, (part of) the encoding of a test case is shown in Figure~\ref{fig:format} where each part of encoding is labeled in the figure. 
It contains 192 bytes, of which the first 96 bytes are initial configuration and the last 96 bytes are a sequence of two function calls and the corresponding input parameters. As there are three string parameters, the first 3 bytes including \texttt{0x05}, \texttt{0x05} and \texttt{0x05} encode the length of \texttt{\_response}, \texttt{\_question} and \texttt{\_answer} respectively. The remaining \texttt{0x05} values are used when there are more than 3 dynamic variables.

Before executing the test case, the bit vector is decoded to a test case according to our internally defined protocol.
Note that the bits in the bit vector may be correlated with each other in multiple ways. For instance, the bits presenting the length of a variable-length value must be equal to the `length' of the value.\\

\noindent \emph{Fitness} After executing the $seeds$ at line 4 in Algorithm~\ref{fig:algo}, function $fitToSurvive()$ is called to evaluate the fitness of the $seeds$ according to a fitness function. Note that the fitness function plays an extremely important role.

In sFuzz, we combine two complementary strategies. One is adopted from AFL, which works as follows. While $seeds$ are executed, sFuzz monitors the execution and records the branches that each test case cover. A test case is deemed `fit to survive' if it covers a new branch in the contract, e.g., a branch which is not covered by any test case in $suite$. This strategy has been shown to be effective in many settings~\cite{afl} and indeed our experimental results show that it is effective in covering most of the branches (see Section~\ref{sec:evaluation}).

Although the AFL strategy allows us to quickly cover most of the branches, it often makes very slow progress in covering the remaining ones afterwards, i.e., often those branches which are with strict conditions. The reason is that most likely the randomly generated test cases would fail to satisfy the strict condition. In such a case, the above fitness function offers little feedback and guideline on how to generate new test cases. For instance, the probability of satisfying the second condition at line 10 of Figure~\ref{examplesol} is as low as $\frac{1}{2^{256}}$ (if we assume that every value is equally likely to be generated). Intuitively, however, it is clear that a test input with $msg.value=200$ is `closer' to satisfy the condition than a test input with $msg.value=10000000$. sFuzz thus integrates an adaptive strategy which selects $seeds$ based on a quantitative measure on how far a test case is from covering any just-missed branch.

Let $br_n$ be a just-missed branch in $\cal S$, i.e., an uncovered outgoing edge from a branching node $n$ in $\cal S$ and $n$ has been covered. The idea is to define a function $distance(t, br_n)$ where $t$ is a test case to return a quantitative measure on how far the branch $br_n$ is from being covered by $t$.

Assume that $br_n$ is labeled with a condition $c$. Note that $c$ can be either $true$, $false$, $a==b$, $a~\text{!=}~b$, $a>=b$, $a>b$, $a<=b$, or $a<b$ at the byte-code level where $a$ and $b$ are variables or constants. In our setting, since $br_n$ is assumed to be a just-missed branch, $c$ must not be $true$ (otherwise $br_n$ must be covered already). Function $distance(t, br_n)$ is then defined as follows.
\[
    distance(t, br_n)=\left\{
                \begin{array}{ll}
                  K & \mbox{if $c$ is $false$}\\
                  \mid a-b \mid +~K & \mbox{if $c$ is $a==b$}\\
                  K & \mbox{if $c$ is $a~\text{!=}~b$}\\
                  b-a+K & \mbox{if $c$ is $a>=b$ or $a>b$}\\
                a-b+K & \mbox{if $c$ is $a<=b$ or $a<b$}\\
                \end{array}
              \right.
\]
where $K$ is a constant which represents the minimum distance. It is set to be 1 in sFuzz. Intuitively, $distance(t, br_n)$ is defined such that the closer the branch is from being covered, the smaller the resultant value is.

\begin{algorithm}[t]
\small
\SetAlgoVlined
let $newSeeds$ be an empty set of test cases\;
\ForEach{$seed$ in $seeds$}{
    \If{$seed$ covers a new branch}{
        add $seed$ into $newSeeds$\;
    }
}
\ForEach{uncovered branches $br_n$}{
    let $min$ be $+\infty$; let $t$ be a dummy test case\;
    \ForEach{$seed$ in $seeds$}{
        \If{$distance(t, br_n) < min$}{
            let $min$ be $dist(t, br_n)$\;
            let $t$ be $seed$\;
        }
    }
    add $t$ into $newSeeds$\;
}
return $newSeeds$\;
\caption{Algorithm $fitToSurvive(seeds)$}
\label{fig:algo2}
\end{algorithm}

With the above, function $fitToSurvive(seeds)$ then selects the seeds as shown in Algorithm~\ref{fig:algo2}. The loop from line 2 to 4 goes through every test case to select those which cover a new branch. Afterwards, for each just-missed branch $br_n$ in the smart contract, the loop from line 5 to line 11 selects a test case from $seeds$ which is the closest to cover the branch according to $distance(t, br_n)$. Note that one seed is selected for each just-missed branch, which makes this algorithm a lightweight multi-objective optimization approach. All selected seeds are then used for crossover and mutation to generate more test cases in the next step. We refer the readers to Section~\ref{example} for an example. \\

\noindent \emph{Remark} The above-described strategy is inspired by search-based software testing (SBST)~\cite{mcminn2004search,harman2010theoretical} and yet it differs from SBST in several ways. The high-level reason for the difference is that having an AFL-based approach for fuzzing requires us to run test cases efficiently whereas existing SBST's seed selection strategy is time-consuming. Furthermore, due to the stack-based implementation of EVM, implementing existing the SBST strategy is infeasible. In the following, we present the differences in detail.

\begin{figure}[t]
    \centering
    \includegraphics[width=0.45\linewidth]{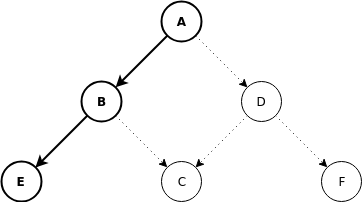}
    \caption{A control flow graph}
    \label{fig:appr_dist}
\end{figure}

First, existing state-of-the-art SBST techniques (i.e., the one in EvoSuite~\cite{harman2010theoretical}) measures how far a test case $t$ is from covering any uncovered branch (not only those just-missed ones) in a more complicated way. That is, given CFG ${\cal S} = (N, i, E)$, let the distance from a node $n_1$ to node $n_2$ to be the minimum number of edges along any path from $n_1$ to $n_2$. 
Let $br_n$ be any uncovered branch and $m$ be a node covered by $t$ which is the nearest node to $n$, i.e., $m$ has a minimum distance to $n$ compared to any other node covered by $t$. SBST uses the following function to measure how far $t$ is from covering $br_n$. 
\[
    dist(t, br_n) = appr\_dist(t, br_n) + norm(distance(t, br_m))
\]
where $br_m$ is an outgoing edge of $m$ which is along the shortest path from $m$ to $n$. Note that if $m$ is $n$ (i.e., in case $br_n$ is just-missed), $br_m$ is simply $br_n$. Function $appr\_dist(t, br_n)$ is a measurement of how far branch $br_n$ is from being covered by test case $t$, i.e., the distance from $m$ to $n$ plus 1. For instance, given a control flow graph as in Figure \ref{fig:appr_dist}, if $t$ covers only the edge $A \rightarrow B \rightarrow E$, $appr\_dist(t, C) = 1$ since there is one branch from $B$ to reach $C$ and there are two branches from $A$ to reach $C$ via $D$. Similarly, $appr\_dist(t, F) = 2$. Lastly, function $norm(x)$ is a normalization function which normalizes the results of $distance(t, br_m)$ to a value between 0 and 1.
One such function is $norm(x) = 1 - 1.001^{-\mid x \mid }$~\cite{harman2010theoretical}.

Applying the above strategy in fuzzing Solidity smart contracts is inefficient, if not infeasible, for multiple reasons. First, calculating $appr\_dist(t, br_n)$ would require us to construct the complete CFG. However, constructing the CFG based on bytecode only is highly nontrivial. In EVM, branches are realized with the opcode \emph{jumpi}, with a value representing the target program counter dynamically at runtime. The only way to know the target is to fully simulate the stack, which is expensive. Second, even if we have the CFG, computing $appr\_dist(t, br_n)$ is still expensive. Given a CFG with $K$ uncovered nodes. To maintain a list of `best' test cases for each uncovered node, we have to calculate $appr\_dist(t, br_n)$ for all $K$ uncovered nodes, i.e., by building a table of the shortest paths from all nodes to these $K$ nodes. Furthermore, whenever a new node is covered, $appr\_dist(t, br_n)$ must be updated. The overhead is unreasonable given that efficiency is key for AFL-based fuzzing. By focusing on just-missed branches, sFuzz avoids both problems. That is, $appr\_dist(t, br_n)$ is always 1 for any just-missed branch $br_n$ since node $n$ must have been covered. Furthermore, because it is constant for any uncovered branch, we can simply skip it in $dist(t, br_n)$ and so that $dist(t, br_n)$ is reduced to $distance(t, br_n)$, without even the need to normalize. This further reduces the overhead.

Another key difference between sFuzz's strategy and existing SBST's is the multi-objective searching strategy.
The multi-objective search strategies in existing SBST consider each uncovered branch as an objective and select Pareto-optimal seeds to evolve in next generation. Given a set of uncovered branch $\{b_1, b_2, ..., b_m\}$, a set of seeds $\{t_1, t_2, ..., t_n\}$, we say $t_i$ is more Pareto-optimal than $t_j$ if $\forall k\in 0..m$, $distance(t_i, b_k) < distance(t_j, b_k)$.
Otherwise, we say that $t_i$ and $t_j$ are Pareto-equivalent. All Pareto-equivalent seeds form a Pareto frontier and the seeds can fall into several Pareto frontiers.
Existing SBST selects the most Pareto-optimal seeds to evolve. A known problem for such a strategy~\cite{dyanmosa} is that the number of seeds in the same Pareto frontier soars with the increase of the number of objectives (i.e., uncovered branches). For example, there could be hundreds of seeds in the most Pareto-optimal frontier with only 3-5 objectives, which makes it hard to select the most promising seeds and increases the runtime overhead.
In contrast, sFuzz keeps one best seed for each just-missed branch (line 6--11 in Algorithm~\ref{fig:algo2}) and as a result, the number of seeds remains small (i.e., equivalent to the number of just-missed branches). 
Our experimental results show that such a strategy balances effectiveness in identifying good seeds and efficiency well.

\subsection{Crossover and Mutation} \label{deter}
Function $crossoverMutation()$ generates new test cases based on those in $seeds$ through crossover and mutation.
sFuzz adopts all of the crossover strategies from AFL and introduces news ones specific for smart contracts. Furthermore, due to correlation between parameters of a test case, sFuzz additionally makes sure the generated test cases are valid. For instance, sFuzz (1) randomly chooses two test cases from $seeds$; (2) breaks the two test cases into two pieces at a selected position; and (3) swaps the second pieces to form two new test cases. Note that due to correlations between the bits representing a test case, there is no guarantee that the resultant test cases are valid and thus sFuzz always checks for validity and discard those invalid ones.

\begin{table}[t]
    \centering
    \footnotesize
    \caption{Mutations for fix-length values}
    \begin{tabular}{|l|}
        \hline
        Name \\ 
        \hline
        \emph{pruneMethodCall} (new) \\ 
        \hline
        \emph{addMethodCall} (new) \\ 
        \hline
        \emph{swapMethodCall} (new) \\ 
        \hline
        \hline
        \emph{singleWalkingBit}, \emph{twoWalkingBit}, \emph{fourWalkingBit} \\ 
        1/2/4 consecutive bits \\
        \hline
        \emph{singleWalkingByte}, \emph{twoWalkingByte}, \emph{fourWalkingByte} \\ 
        \hline
        \emph{singleArith}, \emph{twoArith}, \emph{fourArith} \\ 
        \hline
        \emph{singleInterest}, \emph{twoInterest}, \emph{fourInterest} \\ 
        \hline
        \emph{overwriteWithDictionary} \\ 
        \hline
        \emph{overwriteWithAddressDictionary} \\ 
        \hline
    \end{tabular}
    \label{tab:mutationoperators}
\end{table}
Mutation is another way of generating new test cases. Given a \emph{seed} encoded in the form of a bit vector, sFuzz supports a 
set of mutation operators to generate new test cases. All mutation operators are shown in Table~\ref{tab:mutationoperators}.

Recall that a test case is in the form of an initial configuration and a sequence of function calls with concrete parameters. The first three mutation operators aim to alter the sequence of function calls, by pruning a function call, adding a function call or swapping two function calls. When a function call is pruned (or added or swapped), the corresponding concrete parameters are pruned (or added or swapped) accordingly.

For those values in a test case other than those representing the called functions, sFuzz categorizes them into two groups. The first group contains those values which have fixed-length (e.g., a parameter of type $uint256$). sFuzz systematically applies the remaining mutation operators shown in Table~\ref{tab:mutationoperators} to generate new values, which are inspired by the mutation operators in AFL. Note that account addresses (and balances) are handled slightly differently (refer to the last row in the table) as there are special format requirements. Each address has 32 bytes, in which the last 20 bytes contain the address value and the first 12 bytes contain the balance of the address. For instance, the value \texttt{0xff00...00...00f0} represent an address \texttt{0xf0} with balance \texttt{0xff0000000000000000000000}.

The second group contains those values which have variable-length (e.g., a parameter of type $array$). For such values, their lengths are encoded as part of the test case as well. We thus first mutate the value representing the length in such a way that the result is a random value between 0 and 255 where 255 is an upper bound. If the new length is less than the current one, the corresponding value is shortened accordingly by pruning the additional bits. If the length is more than the current one, random type-compatible values are padded accordingly. 

Note that we discard identical test cases generated through either crossover or mutation. Furthermore, although we do not set a limit on the number of mutations generated from a test case, we apply multiple heuristics adopted from AFL to reduce the number of mutations. For instance, if applying the \emph{WalkingByte} mutation to a block of 32 bytes does not result in any test case which covers a new branch, in the next stages sFuzz will not mutate that block. We refer the readers to AFL for details on these heuristics~\cite{afl}.




 \label{approach}
\section{Implementation} \label{sec:implementation}
sFuzz is implemented in C++ with an estimated 4347 lines of code. It is publically available (https://sfuzz.github.io). It has 3 main components: \textit{runner}, \textit{libfuzzer} and \textit{liboracles}. \\

\noindent {Component \textit{runner}} manages the execution of the test cases. sFuzz takes as input the bytecode of a smart contract along with the ABI (i.e., application binary interface, which can be generated automatically using existing tools) of the contract. The \textit{runner} then generates a bash script file which contains a list of commands to analyze the ABI, and set options for the other two components.

The \textit{runner} sets up a 
test network based on which smart contracts are deployed and transactions are executed. To generate test cases for functions with address-type parameters, sFuzz deploys a pool of externally owned accounts in the 
test network with random balances. The pool size is less than or equal to the number of address-type parameters because it is possible to set the same address to multiple address-type parameters.
The values for address-type parameters are then chosen randomly from this pool. In addition, sFuzz deploys two special smart contracts as attackers, i.e., a \textit{normal attacker} and a \textit{reentrancy attacker}. Each attacker is set as the owner of the contract under test in turn. The \textit{normal attacker} throws an exception whenever other contracts call its payable fallback function. The \textit{reentrancy attacker} calls back the function which makes a call to its payable fallback function. If the attacker fails to call back, it acts as a \textit{normal attacker}. Note that the \textit{reentrancy attacker} is only loaded to detect \textit{Reentrancy} vulnerability.
Otherwise, the \textit{normal attacker} is loaded to avoid call loops of \textit{Reentrancy Attacker} which significantly reduces the speed of sFuzz.
\\

\noindent{Component \textit{libfuzzer}} solves the test generation problem, i.e., how to selectively generate test cases, by implementing the fuzzing strategy presented in the previous sections. It is responsible for multiple tasks.

First, it constructs the CFG of the given smart contract on-the-fly. Ideally, we would like to construct the CFG statically before fuzzing. However, constructing the CFG based on bytecode only is highly nontrivial. In EVM, branches are realized with the opcode \emph{jumpi}, with a value representing the target program counter dynamically at runtime. The only way to know the target is to fully simulate the stack, which is expensive. Therefore, sFuzz constructs the CFG on-the-fly while fuzzing. That is, whenever the opcode \emph{jumpi} is executed, the two destinations are recorded. If these two destinations are not part of the CFG yet, two new nodes are created accordingly representing the two destinations in the CFG.

Second, component \textit{libfuzzer} implements the fuzzing algorithm discussed in Section~\ref{sec:fuzz/smart/contracts}. One optimization is that we
identify \emph{view} functions (i.e., those which do not change any variables) and exclude them from test case generation. The justification is that these view functions do not change the states and having them does not additionaly expose those vulnerabilities sFuzz targets at (see below). Note that view functions are marked by \texttt{view}, \texttt{pure} or \texttt{constant} keywords, sFuzz reads ABI file to recognize them.\\


\noindent {Component \textit{liboracles}} solves the oracle problem, i.e., it monitors the execution of a test case and checks whether there is a vulnerability according to an extensible library of oracles used in sFuzz. sFuzz monitors the execution of test cases through the hooking mechanism supported by EVM. Whenever EVM executes an opcode, it creates an event containing read-only execution information, such as the values of the stack, memory, program counter, and the current executed opcode. sFuzz monitors these events for constructing the CFG and computing $distance(t, br_n)$, as well as logs the events for vulnerability detection. To reduce the execution overhead, vulnerability detection is conducted offline in batches (i.e., once for every 500 test cases). This design allows sFuzz to easily support different versions of Solidity, i.e., by simply replacing the EVM packed in sFuzz.

sFuzz has an extensible architecture which allows it to easily support different oracles as well. Currently, sFuzz supports 8 oracles inspired by the previous work~\cite{Jiang:2018:CFS:3238147.3238177,luu2016making}.
Since these oracles are not our main contribution, we refer the readers to~\cite{Jiang:2018:CFS:3238147.3238177,luu2016making} for details.

These oracles are checked 
based on the logs of test cases. For instance, to check if a test case expose the \emph{Gasless Send} vulnerability, we check that whether test case executes a \emph{CALL} instruction with some data greater than 0 when the gas is equal to 2300.
The test cases that expose vulnerabilities in the contract are kept in a separate test suite and reported to the user together with the vulnerabilities that they expose. Note that by design, sFuzz always reports true positives 
according to our definition of 
vulnerability except in the case of \emph{Freezing Ether}. However, in practice, a reported vulnerability might be a false positive as it may be what the user intended (i.e., our definition of vulnerability is too strict). In the case of \emph{Freezing Ether}, the identified `warning' might be a false positive if there exist some test cases which call $send()$ or $transfer()$ but such test cases are never generated. Technically, the problem of checking whether there is \emph{Freezing Ether} vulnerability can only be solved if we cover all feasible opcode (which is often infeasible).
 \label{implementation}
\section{Experiments and Evaluation}  \label{sec:evaluation}
In this section, we evaluate sFuzz through multiple experiments. The experiments are designed to answer the following research questions (RQ).
\begin{itemize}
    \item \emph{RQ1: How efficient is sFuzz?}
    \item \emph{RQ2: Is sFuzz effective in finding smart contract vulnerabilities and obtaining high code coverage?} 
    \item \emph{RQ3: Is the adaptive strategy useful?}
\end{itemize}
Our test subjects include
4112 smart contracts which we collect from EtherScan~\cite{etherscan}.
These contracts are implemented using Solidity 4.2.24, which is the most popular version of Solidity.
Moreover, the source code for these contracts are available, which makes the evaluation more accurate.
We note that sFuzz can run with bytecode only.
For a baseline comparison, we compare sFuzz with a fuzzer named ContractFuzzer reported in~\cite{DBLP:conf/pldi/GodefroidKS05} and a symbolic execution tool named Oyente reported in~\cite{luu2016making}.
Other fuzzers for smart contracts have been mentioned in~\cite{DBLP:conf/uss/KruppR18}. However, we fail to find the reported tools online or through the authors. We run the experiments 3 times and report the average as the result. All experimental results reported below are obtained on an Ubuntu 18.04.1 LTS machine with Intel Core i7 and 16GB of memory. We use the default initial configuration as presented in Section~\ref{process}.


\subsection{Efficiency} \label{effi}
To answer RQ1, we systematically apply sFuzz, ContractFuzzer and Oyente on all 
4112 smart contracts. 
To save time, each contract is run for 2 minute in this experiment. Note that in general the adaptive fuzzing strategy takes time to show its effectiveness (as we will show later) and thus this setting gives an edge to other tools.

\begin{figure}[t]
    \centering
    \includegraphics[width=\linewidth]{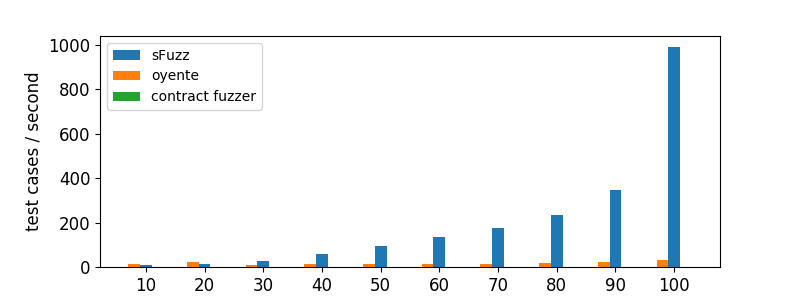}
    \caption{Efficiency comparison between sFuzz, Oyente, and ContractFuzzer}
    \label{fig:speed}
\end{figure}

We measure the efficiency of sFuzz by counting how many test cases are generated and executed per second. Naturally, a test case for a more complicated contract (e.g., with many loop iterations) takes more time to execute. Thus, we show how efficiency varies for different contracts.
Figure~\ref{fig:speed}
summarizes the result, where each bar represents 10\% 
(about 400)
of the fuzzed contracts and the y-axis shows the number of test cases generated and executed per second. The contracts are sorted according to how efficiently it can be fuzzed. From the figure, we observe that the efficiency varies significantly over different contracts, i.e., sFuzz generates and executes more than 989 test cases per second on average for the top 10\% of the contracts, and less than 14 test cases for the bottom 20\%. On average, sFuzz generates and executes more than 208 test cases per second.

Figure~\ref{fig:speed} also compares the efficiency of sFuzz with Oyente and ContractFuzzer.
From the results, we observe that sFuzz is significantly more efficient than other tools. On average, ContractFuzzer and Oyente generate and execute 0.1 and 16 test cases per second respectively.
There are multiple reasons why sFuzz is much faster. First, ContractFuzzer simulates the whole network and manages the blockchain (e.g., commit state changes to storage and append new mined blocks to blockchain after function calls), whereas sFuzz simulates only details of network or blockchain which are relevant to vulnerabilities in smart contracts. Second, sFuzz has a highly optimized implementation in C++, whereas ContractFuzzer is based on Node.js and Go language. 
In the case of Oyente, because it is a symbolic execution tool, Oyente is expected to run slower than a fuzzer like sFuzz.



We further conduct an experiment to measure the overhead of monitoring the execution of a test case (using the hooking mechanism) and the overall overhead of the fuzzing process (including the overall of monitoring the execution, constructing the CFG, mutating the test cases and comparing them, etc.). We apply sFuzz to a set of 60 randomly selected contracts and measure the time spent on executing the test cases, monitoring the execution and other steps of the fuzzing process. The results show that on average the monitoring consumes about 10\% of the total execution time and the overhead of the fuzzing process (including monitoring) is about 14\%. This is very efficient compared to the reported overhead in other fuzzers~\cite{yun2018qsym}.


\subsection{Effectiveness}
\begin{figure}[t]
    \centering
    \includegraphics[width=\linewidth]{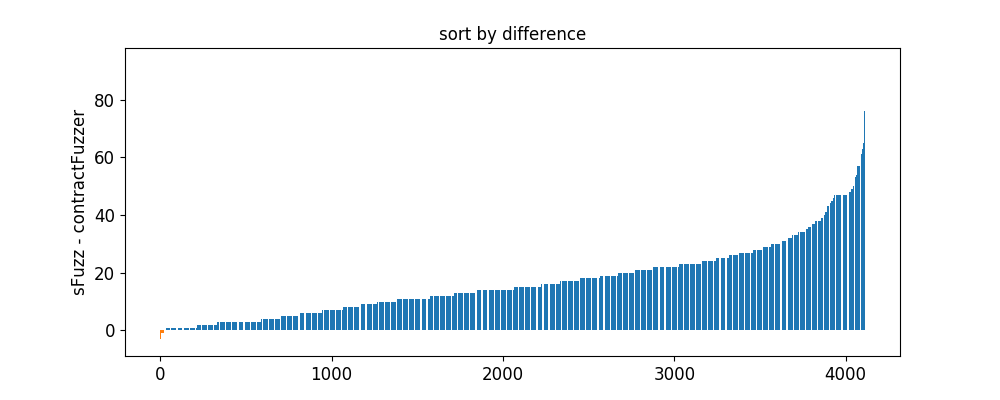}
    \caption{Coverage comparison between sFuzz and ContractFuzzer} 
    \label{fig:coverage1}
\end{figure}

\begin{figure}[t]
    \centering
    \includegraphics[width=\linewidth]{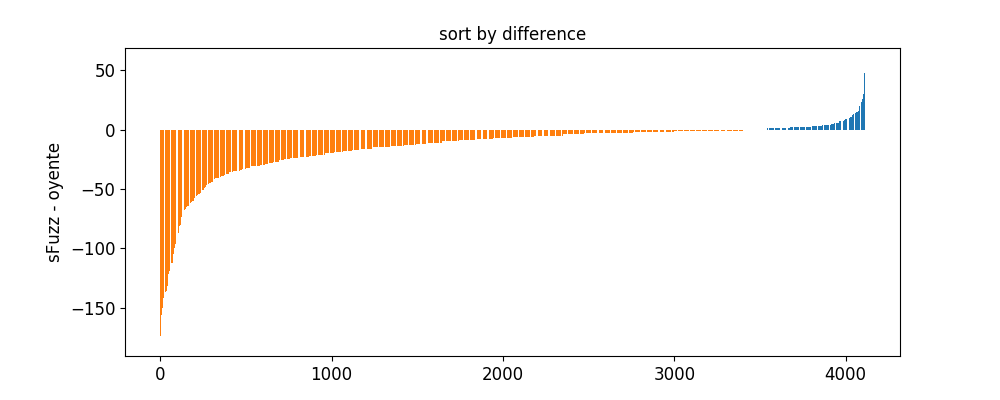}
    \caption{Coverage comparison between sFuzz and Oyente} 
    \label{fig:coverage2}
\end{figure}

To answer RQ2, we aim to measure the branch coverage achieved by the test suite generated for each smart contract, as well as count the number of vulnerabilities identified. However, measuring branch coverage precisely is highly non-trivial due to, for instance, the problem of infeasible branches. Thus, we instead measure the number of distinct branches covered by the generated test suite. Figure~\ref{fig:coverage1} summarizes a comparison between sFuzz and ContractFuzzer in terms of the number of distinct branches covered. The $y$-axis is the number of branches covered by sFuzz minus that of ContractFuzzer and each point on the $x$-axis represents a smart contract. The contracts are sorted by their $y$-axis value.
Similarly, Figure~\ref{fig:coverage2} shows the comparison between sFuzz and Oyente.

For most of the smart contracts 
(i.e., 4077 of 4112 contracts)
sFuzz covers more branches than ContractFuzzer. 
To our surprise, ContractFuzzer managed to cover more branches for 35 contracts. A closer investigation shows that the number of branches covered by ContractFuzzer is inflated for the following reasons. First, as sFuzz does not execute \textit{view} functions (for efficiency reasons), all branches in these functions are not counted. Because \textit{view} functions do not modify the state of a smart contract, they are considered irrelevant to vulnerabilities.
Second, ContractFuzzer sometimes generates invalid test cases which fail mandatory constraints and cover additional branches.
Mandatory constraints are generated by the compiler (i.e., the Solidity compiler) and are embedded in the bytecode to assert the correctness logic of function calls or data types. For example, ContractFuzzer invokes a \textit{fallback} function of a non-fallback contract or sends Ethereum to functions which are not marked with the \textit{payable} keyword. As a result, the mandatory constraints are failed which lead to branches which signal an error in the test case being covered.

In the case of Oyente, in 3402 contracts, Oyente covers more branches than sFuzz. An investigation shows that Oyente analyzes every function separately and thus has to assume that state variables can take arbitrary values (without considering their initial values or constraints on how the values are updated). As a result, Oyente can easily satisfy almost all conditions in smart contracts. Given the sample contract A in Figure \ref{examplesol3}, Oyente covers 99.1\% EVM code and discovers an integer overflow vulnerability. It means that these conditions: $id == 9$ and $balances[msg.sender] > 10$ are satisfied. However, it is impossible as there is no way to change values of $id$ and $balances[msg.sender]$. Often, a condition in smart contract is the comparison between local/parameter variables and state variables, e.g., $balances[msg.sender] > value$ (whether sender has enough Ethereum to deduce). In such cases, sFuzz must call the function which sets certain values to the state variables before satisfying them whereas Oyente assigns arbitrary values directly to state variables. It is apparent to us that Oyente's approach is flawed and would `cover' many infeasible paths.

\begin{figure}[t]
\centering
\begin{minipage}[h]{0.8\linewidth}
\begin{lstlisting}
contract A {
  mapping(address => uint) balances;
  uint id = 10;
  function main(uint x, uint y) {
    if (id == 9) {
      if (balances[msg.sender] > 10) {
        uint sum = x + y; } } } }
\end{lstlisting}
\end{minipage}
\caption{Oyente visits infeasible branches}
\label{examplesol3}
\end{figure}

In the following, we summarize the number of vulnerable contracts discovered by sFuzz in each category. The results are shown in Table~\ref{table:vulernability}. The first column shows the type of vulnerability. The next three columns show the number of vulnerable contracts found by sFuzz, ContractFuzzer
and Oyente
respectively. The sub-column \# show the number of contracts that have the vulnerability according to each vulnerability type and the second sub-column is the percentage of true positives of the identified vulnerabilities. For all categories, sFuzz finds more vulnerable contracts than ContractFuzzer. Note that ContractFuzzer removes \emph{Freezing Ether} from their source code and does not check \emph{Integer Overflow/Underflow}. 
In total, sFuzz finds vulnerabilities in 
1113 contracts, i.e., 24 times more than that of ContractFuzzer.

To evaluate the soundness of sFuzz, we manually examine the identified vulnerable contracts to check whether they are true positives or not. However, we are unable to manually check all the identified vulnerability for two reasons. First, there is an overwhelming number of vulnerabilities. 
Instead, we randomly sample 50 vulnerable contracts with source code in each category and manually check whether the identified vulnerability is a true positive or not. If there are fewer than 50 vulnerable contracts with source code in the category, we check all of them.

\begin{table}[t]
\centering
\caption{Vulnerabilities} 
\label{table:vulernability}
\addtolength{\tabcolsep}{-2.5pt}
{\footnotesize \begin{tabular}{|l|c|c|c|c|c|c|c|c|c|}
\hline
 \multirow{2}{*}{\textbf{Vulnerability Type}} & \multicolumn{2}{c|}{\textbf{sFuzz}} & \multicolumn{2}{c|}{\textbf{ContractFuzzer}} &
 \multicolumn{2}{c|}{\textbf{Oyente}}
 \\
\cline{2-7}
 & \textbf{\#} & \textbf{true posi.} & \textbf{\#} & \textbf{true posi.} & \textbf{\#} & \textbf{true posi.}\\
\hline
\emph{Gasless Send} & 764 & 100\% & 14 & 100\% & 0 & N.A. \\
\emph{Exception Disorder} & 36 & 100\% & 6 & 100\% & 0 & N.A. \\
\emph{Reentrancy} & 29 & 100\% & 3 & 100\% & 52 & 60\% \\
\emph{Timestamp Dependency} & 243 & 86\% & 28 & 86\% & 102 & 100\% \\
\emph{Block Number Dependency} & 59 & 80\% & 16 & 95\% & 0 & N.A. \\
\emph{Dangerous DelegateCall} & 17 & 100\% & 0 & 100\% & 0 & N.A. \\
\emph{Integer Overflow} & 98 & 100\% & 0 & N.A. & 3350 & 60\% \\
\emph{Integer Underflow} & 224 & 80\% & 0 & N.A. & 2246 & 60\% \\
\emph{Freezing Ether} & 15 & 60\% & 0 & N.A. & 0 & N.A. \\
\hline
\end{tabular}}
\end{table}


For \emph{Gasless Send}, \emph{Exception Disorder} and \emph{Reentrancy} vulnerability, all 50 sampled vulnerable contracts are true positives. 
For \emph{Time-stamp Dependency}, out of the 50 sampled vulnerable contracts, 43 of them are true positives. In the remaining 7 contracts, although  \emph{block.timestamp} and/or \emph{now} is used in a condition, they are irrelevant to the Ether sending part (i.e., no control/data dependency). Rather their values are saved in global variables to record the creation time of specific events. sFuzz mistakenly claims that such cases are vulnerable. For \emph{Block Number Dependency}, 40 out of the 50 sampled vulnerable contracts are true positives. Similarly, the reason for the 10 false positives is the value of \emph{block.number} is assigned to global variables but they are irrelevant to Ether sending process. For \emph{Dangerous DelegateCall}, all 17 sampled contracts are indeed vulnerable. Similarly so for \emph{Integer Overflow}. For \emph{Integer Underflow}, 40 of the 50 identified contracts are indeed vulnerable. The reason for the 10 false positives is because it is non-trivial to identify the correct type of a variable based on bytecode only (e.g., whether it is \emph{uint256} or \emph{uint128}), sFuzz conservatively assumes that all arithmetic operations returning a negative value may be vulnerable. This can be improved by adopting the approach in~\cite{torres2018osiris} to infer types based on EVM bytecode.
Lastly, for \emph{Freezing Ether}, 9 of the 15 identified contracts are true positives. The reason for the 6 false positives is that although there is a program path which allows the contract to send Ether, the program path is not covered and sFuzz falsely assumes that there is no such program path. This percentage of such false positives is expected to be reduced if sFuzz is applied for a longer time (with more branches covered).

\begin{figure}[t]
    \centering
    \includegraphics[width=\linewidth]{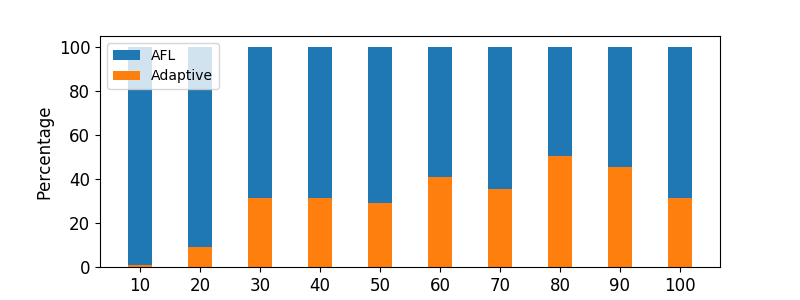}
    \caption{Percentage of test cases due to adaptive strategy}
    \label{fig:adaptiveness-ADA}
\end{figure}

The last column in Table~\ref{table:vulernability} shows the results of Oyente. The results should be taken with a grain of salt since Oyente requires the source code. For instance, it is trivial to know the type of variables with the source code, and thus Oyente identifies many more problems with \emph{Integer Overflow/Underflow}. For the remaining vulnerabilities, Oyente does not support 5 of them; identifies a higher number of vulnerable contracts for \emph{Reentrancy} but with a higher false positive rate; and identifies much fewer vulnerable contracts for \emph{Timestamp Dependency}.

\begin{figure}[t]
    \centering
    \includegraphics[width=\linewidth]{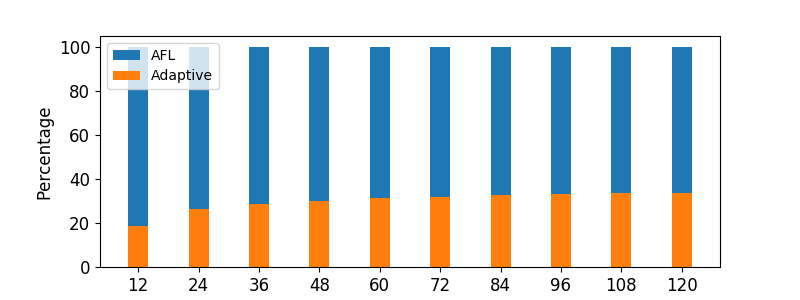}
    \caption{Effective of adaptive strategy over time }
    \label{fig:adaptiveness-AFL}
\end{figure}

\subsection{Adaptiveness}
To answer RQ3, we systematically analyze the test suite generated by sFuzz for each smart contract. Note that each test case covers at least one branch which is not covered by any other test cases. To measure how the two fuzzing strategies implemented in sFuzz complement each other, we count how many test cases in the resultant test suites are generated due to the AFL strategy and how many are due to the adaptive strategy. Note that a test case is judged to be due to the adaptive strategy if and only if it is generated based on a seed selected by line 11 at Algorithm~\ref{fig:algo2}.

The results are shown in Figure~\ref{fig:adaptiveness-ADA}, where the $y$-axis is the percentage of test cases generated by the strategy. Each bar represents 10\% of the contracts. We remark that the two strategies have different targets and thus whether they are effective largely depends on what branching conditions are in the smart contracts. We thus sort the contracts according to the speed of sFuzz.
The bar on the rightmost thus represents the top 10\% contracts. We observe that, as expected, the AFL strategy easily covers most of the branches (since the conditions for executing most branches are not strict). For about 80\% of the smart contracts, the adaptive strategy makes a noticeable contribution, i.e., contributing an average of 31\% of the generated test cases. Given that sFuzz is applied for each contract only for 2 minutes, the result is encouraging as we hypothesize that the effect of the adaptive strategy would be more apparent if sFuzz is applied for a longer period of time.

To test our hypothesis,
we record the percentage of test cases generated by the adaptive strategy every 12 seconds. The results are shown in Figure~\ref{fig:adaptiveness-AFL}, where the $x$-axis is the fuzzing time and each bar shows the percentage after certain number of seconds. We can observe that the percentage of generated test cases by adaptive strategy increases with more fuzzing time. On average, the percentage rises from 18\% after 12 seconds fuzzing to 33\% after 2 minutes fuzzing. From the results, we conclude the adaptive strategy is useful in increasing the coverage of the generated test suites. \\

\noindent
\emph{Threat to validity}
There are both internal threats and external threats to our work. For external threats,
it is probable that sFuzz's performance will vary with the choice of the initial population, as other researchers have noted~\cite{klees2018evaluating}. 
For internal threats, the percentage of true positives in Table~\ref{table:vulernability} may not be accurate as they are approximated by a sample of 50 contracts for each type of vulnerability. In addition, the exact intention of the author of the contract is not always clear, even if we try our best to read the source code.
 \label{evaluation}
\section{Related Work and Conclusion} \label{sec:related}

sFuzz is closely related to existing fuzzers for smart contracts.  ContractFuzzer~\cite{Jiang:2018:CFS:3238147.3238177} is a fuzzer which can check 7 different types of vulnerabilities. Its approach, however, does not use any feedback to improve the test suite.
Echidna~\cite{echidna} is another fuzzer that is reportedly capable of
checking if the contract violates some user-defined properties. However,
we fail to find any publication about it. 

sFuzz is complementary to existing symbolic execution engines for smart contracts. In~\cite{luu2016making}, Luu \emph{et al.} presented an
engine to find potential security bugs in smart contracts.
The tool, however, is neither sound nor complete.
In~\cite{DBLP:conf/uss/KruppR18}, Krupp and Rossow presented teEther,
which is focused on financial transactions
and related vulnerabilities. In~\cite{nikolic2018finding}, Nikolic \emph{et al.} presented
a tool named MAIAN, which 
can find 3 types of trace vulnerabilities.
In~\cite{torres2018osiris}, Torres \emph{et al.} presented Osiris, a tool which combines symbolic execution and taint analysis to discover 3 types of integer bugs in smart contracts.
Different from the above works, sFuzz is a fuzzer and it can be combined with the above engines to form a hybrid fuzzing engine. 

sFuzz is related to work on formal verification of smart contracts. Zeus~\cite{kalra2018zeus} is a framework which verifies the correctness and
fairness of smart contracts based on LLVM.
Bhargavan \emph{et al.} proposed a framework to verify smart contracts formally by transforming
the source code and the bytecode to F*, a language designed for verification~\cite{bhargavan2016formal}.
In~\cite{hirai2016formal}, the author presented an attempt to verify the Deed contract
using Isabelle/HOL~\cite{nipkow2002isabelle}. 

sFuzz is broadly related to work on analyzing smart contracts. In~\cite{delmolino2016step}, Delmolino \emph{et al.} showed that writing a safe smart contract is not a trivial task.
In~\cite{atzei2016survey}, Atzei \emph{et al.} provided a taxonomy for common vulnerabilities in smart contracts with real-world attacks. 
In~\cite{frowis2017code}, the authors performed a call graph analysis
and showed that only 40\% of smart contracts are truthless as their control flows are immutable. 
In~\cite{chen2017under}, Chen \emph{et al.} presented 7 gas-cost programming patterns
and showed that most of the contracts suffer from these gas-cost patterns. 

 \label{related}
To conclude, in this work, we present sFuzz, an adaptive fuzzing engine for EVM smart contracts. Experimental results show that sFuzz is significantly more reliable, faster, and more effective than existing fuzzers. sFuzz is currently under rapid development and has already gained interest from multiple companies and research organizations.  \label{conclusion}

\begin{acks}
This research was supported by the Singapore Ministry of Education (MOE) Acemedic Research Fund (AcRF) Tier 1 grant.
\end{acks}


\bibliographystyle{ACM-Reference-Format}
\bibliography{arf}
\end{document}